\documentclass[12pt]{article}
\usepackage{graphicx}
\newcommand{\be}{\begin{equation}}
\newcommand{\ee}{\end{equation}}
\newcommand{\ba}{\begin{eqnarray}}
\newcommand{\ea}{\end{eqnarray}}

\renewcommand{\kappa}{{k}}

\catcode`\@=11

\@addtoreset{equation}{section}
\def\theequation{\thesection\arabic{equation}}

\def\@normalsize{\@setsize\normalsize{15pt}\xiipt\@xiipt
\abovedisplayskip 14pt plus3pt minus3pt%
\belowdisplayskip \abovedisplayskip
\abovedisplayshortskip  \z@ plus3pt%
\belowdisplayshortskip  7pt plus3.5pt minus0pt}
\def\small{\@setsize\small{13.6pt}\xipt\@xipt
\abovedisplayskip 13pt plus3pt minus3pt%
\belowdisplayskip \abovedisplayskip
\abovedisplayshortskip  \z@ plus3pt%
\belowdisplayshortskip  7pt plus3.5pt minus0pt
\def\@listi{\parsep 4.5pt plus 2pt minus 1pt
            \itemsep \parsep
            \topsep 9pt plus 3pt minus 3pt}}

\def\underline#1{\relax\ifmmode\@@underline#1\else
        $\@@underline{\hbox{#1}}$\relax\fi}
\@twosidetrue \relax

\catcode`@=12

\catcode`\@=11

\def\section{\@startsection{section}{1}{\z@}{3.5ex plus 1ex minus
   .2ex}{2.3ex plus .2ex}{\large\bf}}
\def\thesection{\arabic{section}.}

\def\ps@headings{\def\@oddfoot{}\def\@evenfoot{}
\def\@oddhead{\hbox{}\hfill
        \makebox[.5\textwidth]{\raggedright\ignorespaces --\thepage{}--
        \hfill }}
\def\@evenhead{\@oddhead}
\def\subsectionmark##1{\markboth{##1}{}}
}

\ps@headings

\catcode`\@=12

\relax

\def\figcap{\section*{Figure Captions\markboth
        {FIGURECAPTIONS}{FIGURECAPTIONS}}\list
        {Fig. \arabic{enumi}:\hfill}{\settowidth\labelwidth{Fig. 999:}
        \leftmargin\labelwidth
        \advance\leftmargin\labelsep\usecounter{enumi}}}
 \relax
\def\tablecap{\section*{Table Captions\markboth
        {TABLECAPTIONS}{TABLECAPTIONS}}\list
        {Table \arabic{enumi}:\hfill}{\settowidth\labelwidth{Table 999:}
        \leftmargin\labelwidth
        \advance\leftmargin\labelsep\usecounter{enumi}}}
 \relax
\def\reflist{\section*{References\markboth
        {REFLIST}{REFLIST}}\list
        {[\arabic{enumi}]\hfill}{\settowidth\labelwidth{[999]}
        \Leftmargin\labelwidth
        \advance\leftmargin\labelsep\usecounter{enumi}}}
 \relax

\catcode`\@=11

\def\marginnote#1{}
\newcount\hour
\newcount\minute
\newtoks\amorpm
\hour=\time\divide\hour by60 \minute=\time{\multiply\hour by60
\global\advance\minute by- \hour}
\edef\standardtime{{\ifnum\hour<12 \global\amorpm={am}%
    \else\global\amorpm={pm}\advance\hour by-12 \fi
    \ifnum\hour=0 \hour=12 \fi
    \number\hour:\ifnum\minute<100\fi\number\minute\the\amorpm}}
\edef\militarytime{\number\hour:\ifnum\minute<100\fi\number\minute}
\def\draftlabel#1{{\@bsphack\if@filesw {\let\thepage\relax
  \xdef\@gtempa{\write\@auxout{\string
    \newlabel{#1}{{\@currentlabel}{\thepage}}}}}\@gtempa
    \if@nobreak \ifvmode\nobreak\fi\fi\fi\@esphack}
     \gdef\@eqnlabel{#1}}
\def\@eqnlabel{}
\def\@vacuum{}
\def\draftmarginnote#1{\marginpar{\raggedright\scriptsize\tt#1}}
\def\draft{\oddsidemargin -.5truein
        \def\@oddfoot{\sl preliminary draft \hfil
        \rm\thepage\hfil\sl\today\quad\militarytime}
        \let\@evenfoot\@oddfoot \overfullrule 3pt
        \let\label=\draftlabel
        \let\marginnote=\draftmarginnote

\def\@eqnnum{(\theequation)\rlap{\kern\marginparsep\tt\@eqnlabel}%
\global\let\@eqnlabel\@vacuum}  }
\def\preprint{\twocolumn\sloppy\flushbottom\parindent 1em
        \leftmargini 2em\leftmarginv .5em\leftmarginvi .5em
        \oddsidemargin -.5in    \evensidemargin -.5in
        \columnsep 15mm \footheight 0pt
        \textwidth 250mmin      \topmargin  -.4in
        \headheight 12pt \topskip .4in
        \textheight 175mm
        \footskip 0pt

\def\@oddhead{\thepage\hfil\addtocounter{page}{1}\thepage}
        \let\@evenhead\@oddhead \def\@oddfoot{} \def\@evenfoot{}
}
\def\titlepage{\@restonecolfalse\if@twocolumn\@restonecoltrue\onecolumn
     \else \newpage \fi \thispagestyle{empty}\c@page\z@
        \def\thefootnote{\fnsymbol{footnote}} }
\def\endtitlepage{\if@restonecol\twocolumn \else  \fi
        \def\thefootnote{\arabic{footnote}}
        \setcounter{footnote}{0}}  %\c@footnote\z@ }
\catcode`@=12 \relax
\def\ps@headings{\def\@oddfoot{}\def\@evenfoot{}
\def\@oddhead{\hbox{}\hfill
        \makebox[.5\textwidth]{\raggedright\ignorespaces --\thepage{}--
        \hfill }}
\def\@evenhead{\@oddhead}
\def\subsectionmark##1{\markboth{##1}{}}
}

\ps@headings

\relax

\begin{document}
\begin{titlepage}

\begin{centering}
\begin{flushright}
hep-th/0203241 
\end{flushright}

{\Large {\bf Acceleration of the Universe in Type-$0$ 
Non-Critical Strings }}

\vspace{0.01in}

{\bf G.~A.~Diamandis } and {\bf B.~C.~Georgalas}
\\
{\it Physics Department, Nuclear and Particle Physics Section,
University of Athens,\\ Panepistimioupolis GR 157 71, Ilisia,
Athens, Greece.} \\

{\bf N.~E.~Mavromatos } \\ {\it Department of Physics, Theoretical
Physics, King's College London,\\ Strand, London WC2R 2LS, United
Kingdom.} \\

and \\

{\bf E.~Papantonopoulos} \\ {\it
Department of Physics, National Technical University of Athens,\\
Zografou Campus GR 157 80, Athens, Greece.} \\

\vspace{0.01in}
 {\bf Abstract}

\end{centering}

{\small Presently there is preliminary observational evidence 
that the
cosmological constant might be non zero, and hence that our Universe is 
eternally accelerating (de Sitter). This
poses fundamental problems for string theory, since 
a scattering matrix is not well defined in such
Universes. In a previous paper 
we have presented a model, based on (non-equilibrium) 
non-critical strings, which
is characterized by eventual ``graceful'' 
exit from a de Sitter phase. 
The model is based on a type-0 string theory, involving D3 brane worlds,
whose initial quantum fluctuations induce the non criticality.
We argue in this article 
that this model is
compatible with the current observations.
A crucial r\^ole for the correct ``phenomenology'' of the model 
is played by the relative magnitude of the flux of the 
five form of the type $0$ string to 
the size of five of the extra dimensions, transverse 
to the direction of the flux-field. 
We do not claim, at this stage at least, that this
model is a realistic 
physical model for the Universe, but we find it interesting
that the model cannot be ruled out immediately, 
at least on phenomenological grounds.}

\end{titlepage}

\newpage

\section{Introduction}

Recently there is some preliminary experimental 
evidence
from type Ia supernovae data~\cite{evidomegal},  
which supports the fact that our 
Universe {\it accelerates}
at present: distant supernovae (redshifts  $z \sim 1$) 
data indicate a slower rate of expansion, as compared
with that inferred from data pertaining to nearby supernovae.
Distant supernovae look dimer than they should be, if the 
expansion rate of the Universe would be constant.

This could be a consequence 
of a non-zero 
cosmological constant, which would imply that our Universe would be  
eternally accelerating (de Sitter), according to standard 
cosmology~\cite{carroll}.
Such evidence 
is still far from being confirmed, but it is 
reinforced by combining these data with 
Cosmic Microwave background (CMB) data (first acoustic peak) 
implying 
a 
spatially $\Omega _{\rm total} =1.0 \pm 0.1$ flat 
Universe~\cite{evidenceflat}.

Let us review briefly the current situation.
Specifically, the best fit spatially flat Universe 
to the data of \cite{evidomegal,evidenceflat} 
implies, to $3~\sigma$ confidence level,
that 
\begin{equation} 
\Omega _{M,0} ({\rm matter}) \simeq 0.3~, \qquad {\rm and} \qquad 
\Omega_{X,0} ({\rm dark~energy})
\simeq 1 -\Omega _M \simeq 0.7~, 
\label{bestfit}
\end{equation}
where the subscript $0$ indicate present values.  
If the data have been interpreted right this means 
that $70\%$ of the present energy 
density   
of the Universe consists of an unknown substance 
(``{\it dark energy}'').
For the fit of \cite{evidomegal,evidenceflat} the 
dark energy has been taken to be the standard cosmological constant. 

An important phenomenological parameter, which 
is of particular interest to astrophysicists 
is the {\it deceleration parameter q} of the Universe~\cite{carroll}, 
which is defined as: 
\begin{equation}
q = -\frac{(d^2a_E/dt_E^2)~a_E}{({da_E/dt_E})^2} 
\label{decel}
\end{equation} 
where $a_E(t)$ is the Robertson-Walker scale factor of the Universe, 
and the subscript $E$ denotes quantities 
computed in the so-called Einstein frame, 
that is where the gravity action has the canonical Einstein
form as far as the scalar curvature term is concerned. 
This distinction is relevant in string-inspired effective 
theories with four-dimensional Brans-Dicke type scalars,
such as dilatons, which will be dealing with here. 

It should be mentioned that in standard Robertson-Walker cosmologies
with matter the deceleration parameter can be expressed in terms of the 
matter and vacuum (cosmological. constant) energy densities, $\Omega_M$ and $\Omega_\Lambda$ respectively, as follows:
\begin{equation}\label{decel2} 
           q=\frac{1}{2}\Omega_M - \Omega_\Lambda
\end{equation}
For the best fit 
Universe (\ref{bestfit}) 
one can then infer 
a present deceleration 
parameter 
$q_0 = -0.55 < 0$,  
indicating that the Universe accelerates today. 

If the data have been interpreted right, 
then there are three possible explanations~\cite{carroll}:

\par (i) Einstein's General Relativity is incorrect, and hence 
Friedman's cosmological solution as well. This is unlikely,
given the success of General Relativity and of the 
Standard Cosmological Model in explaining a plethora of other
issues. 

\par (ii) the `observed' dark energy and the acceleration of the Universe 
are due to an `honest' cosmological {\it constant} $\Lambda$ in 
Einstein-Friedman-Robertson-Walker cosmological 
model. This is the case of the best fit Universe (\ref{bestfit})
which matches the supernova and CMB data. 
In that case one is facing the 
problem of eternal acceleration, for the following reason: 
let $\rho_M \propto a^{-3}$ the matter density in the Universe,
with $a (t)$ the Robertson-Walker scale factor, and $t$ the 
cosmological observer (co-moving) frame time. The vacuum energy 
density, due to $\Lambda$, is assumed to be constant in time,
$\rho_\Lambda $ =const. Hence in conventional 
Friedmann cosmologies one has: 
\ba
\Omega_\Lambda /\Omega_M = 
\rho_\Lambda / \rho_M \propto a(t)^{3}~, 
\label{conventional}
\ea
and hence eventually
the vacuum energy density component will dominate over matter.
{}From Friedman's equations then, one observes that the Universe 
will eventually enter a de-Sitter phase, 
in which $a(t) \sim e^{\sqrt{\frac{8\pi G_N}{3}\Lambda}t}$, 
where $G_N$ is the gravitational (Newton's) constant. 
This implies eternal expansion and acceleration, and most importantly
the presence of a {\it cosmological horizon} 
\begin{equation}
  \delta = a(t) \int_{t_0}^\infty \frac{c dt}{a(t)} < \infty 
\label{cosmhorizon} 
\end{equation}
It is this last feature in de Sitter Universes that 
presents problems in defining proper {\it asymptotic states}, and thus 
a consistent scattering matrix
for field theory in such backgrounds~\cite{eternal}. 
The analogy of such global horizons
with microscopic or macroscopic black hole horizons in this respect
is evident, the important physical difference, however, being that 
in the cosmological de Sitter case the observer lives ``inside'' the horizon,
in contrast to the black hole case. 

Such eternal-acceleration Universes are, therefore, 
bad news for critical string 
theory~\cite{eternal}, 
due to the fact that strings are by definition theories
of on-shell $S$-matrix and hence, as such, can only accommodate backgrounds
consistent with the existence of the latter.

\par (iii) the `observed effects' are due to the existence of a 
{\it quintessence} field $\varphi$, which has not yet relaxed in its
absolute minimum (ground state), given that the relaxation time 
is longer than the age of the Universe. Thus we are still in a 
non-equilibrium situation, relaxing to equilibrium gradually.
In this drastic explanation, the vacuum energy density, due to the 
potential of the field $\varphi$ will be time-dependent. 
In fact the data point towards a $1/t^2$ relaxation, with $t \ge 10^{60}$
in Planck units, where the latter number represents the age of the 
observed Universe.

It is this third possibility that we have attempted to adopt
in a proper non-critical string theory framework 
in ref. \cite{dgmpp}. 
Non-critical strings can be viewed as non-equilibrium 
systems in string theory~\cite{emn}. 
The advantage of this non-equilibrium
situation lies on the possibility of an eventual exit
from the de Sitter phase, which would allow proper 
definition of a field-theory scattering matrix, thus avoiding the 
problem of eternal horizons mentioned previously.
This would be a welcome fact from the point of view of string theory.

However, 
it should be stressed that our studies, although encouraging,
are far from being complete from a mathematical point of view. 
Non-critical (Liouville) string theory, with time-like backgrounds,
like the ones used in \cite{dgmpp},
is not completely understood at present~\cite{ddk,aben}, although
significant progress has been made towards this direction. 
The main reason for this is the fact that world-sheet
Liouville correlation functions,
do not admit a
clear interpretation as ordinary scattering ($S$) matrix on-shell 
amplitudes in target space. Rather, they 
are \$ -matrix amplitudes, connecting asymptotic density matrices, as  
appropriate for 
the `open system' character of such string theories~\cite{emn,emnsmatrix}. 
Such systems may thus be of relevance to cosmology, 
especially after the above mentioned 
recent experimental evidence
on a current era acceleration of the Universe.

Exit from de Sitter inflationary phases is another feature 
that cannot be accommodated (at least to date)
within the context of critical strings~\cite{emn}. 
This is mainly due to the fact that
such a possibility requires time-dependent backgrounds in string theory
which are also not well understood. 
On the other hand there is sufficient evidence that such 
a `graceful exit possibility' from the inflationary 
phase can be realized in non-critical
strings, with a time-like signature of the Liouville field, which thus 
plays the r\^ole of a Robertson-Walker comoving-frame time~\cite{emn}.
The evidence came first from toy two-dimensional specific 
models~\cite{grace}, 
and recently was extended~\cite{dgmpp} to four-dimensional models 
based on the so-called type-0 non-supersymmetric strings~\cite{type0}.
The latter string theory has four-dimensional brane worlds, whose fluctuations 
have been argued in \cite{dgmpp} to lead to super criticality
of the underlying string theory, necessitating Liouville dressing
with a Liouville mode of time-like signature. In general, Liouville 
strings become critical strings after such a dressing procedure,
but in one target-space dimension higher. 
However in our approach~\cite{emn,dgmpp}, instead of increasing the 
initial number of target space dimensions ($d=10$ for type-0 strings), 
we have 
identified the world-sheet zero mode of the Liouville field with the 
(existing) target time. In this way, the time may be thought of as 
being responsible of re-adjusting itself (in a non-linear way), 
once the fluctuations
in the brane worlds occur, so as to restore the conformal invariance
of the underlying world sheet theory, which has been disturbed by the 
brane fluctuations. 

One of the most important results of \cite{dgmpp} is the 
appearance of a time-dependent central charge deficit 
in the underlying conformal theory, acting as a vacuum energy density
in the respective target-space lagrangian. This also is crucial for 
a `graceful 
exit' from the inflationary de Sitter phase, and the absence of 
eternal acceleration. In fact the asymptotic (in time) theory 
is that of a flat (Minkowski) target-space $\sigma$-model with a 
linear dilaton~\cite{aben} in the string frame. 
In the Einstein frame, i.e. in a redefined
metric background in which the Einstein curvature term in the 
target-space effective action has the canonical normalization, 
the universe is linearly expanding, which is the limiting
case in which the horizon (\ref{cosmhorizon}) diverges logarithmically;
hence such a theory can admit properly defined asymptotic states 
and $S$-matrix amplitudes. The linear dilaton background has 
been shown~\cite{aben} to be a consistent background for string theory,
despite being time dependent,
in the sense of satisfying factorizability (for certain discrete values
of the asymptotic central charge though), modular invariance and 
unitarity.

Another important aspect of our solution 
is the fact that the extra bulk dimensions are compactified
in such a way that one is significantly larger than the others, thereby
leading to effective five-dimensional brane world scenaria. 
The reason for this is appropriately chosen five-form flux background.
This feature is, we believe, one of the most important
ones of the type-0 stringy cosmologies, which are known to be characterized
by the existence of non-trivial flux form fields coming from the 
Ramond sector of the brane worlds~\cite{type0,dgmpp}.

The reader might object to our use of type-0 backgrounds due to 
the existence of tachyonic backgrounds. 
Although at tree level it has been demonstrated that 
the above-described flux forms can stabilized such backgrounds,
by shifting away the tachyonic mass poles, however, recently this
feature has been questioned at string loop level.
Nevertheless, in the context of our cosmological model,
such quantum instabilities are expected probably as a result
of the {\it non-equilibrium} nature of our 
relaxing background, in which the time dependent dilaton field
plays the r\^ole of the quintessence field.
Indeed, as demonstrated in \cite{dgmpp} the 
asymptotic in time value of the tachyon background is zero,
and hence such a field disappears eventually from the spectrum,
which is 
consistent with the asymptotic equilibrium nature of the 
ground state.

The structure of the present article,  
is as follows: in section 2 we review 
the main features of the cosmological model of \cite{dgmpp}. 
In section 3 we discuss the phenomenology of the model as implied 
by the current astrophysical observations. In particular, we 
demonstrate that the present-era values of the 
deceleration parameter of the type-0 non-critical string 
Universe, the ``vacuum energy density'', 
and the Hubble parameter all match the experimental data. 
A crucial r\^ole for this, in particular for matching the 
order-one value of the deceleration parameter, is played
by the relative magnitude of the flux of the five form
field of the type 0 string to the size of the five extra dimensions
transverse to the direction of the flux. 
Conclusions are presented in section 4. 

\section{Cosmology with type-0 non-critical Strings: A brief review} 

In this section we review the main features of the Cosmological 
String model of \cite{dgmpp}. 
We commence our discussion by recalling that 
the effective ten-dimensional target space action
of the type-$0$ String, 
to ${\cal O}(\alpha ')$ in
the Regge slope $\alpha '$, assumes the form~\cite{type0}:
 \begin{eqnarray}\label{actiontype0} S&=&\int d^{10} x
\sqrt{-G}\Big{ [}e^{-2\Phi}\Big{(}R + 4(\partial_M \Phi)^2
-\frac{1}{4}(\partial_M T)^2 - \frac{1}{4}m^2T^2 \nonumber \\&~~ &
-\frac{1}{12}H_{MNP}^2\Big{)} -  \frac{1}{4}(1 + T +
\frac{T^2}{2})|{\cal F}_{MNP\Sigma T}|^2\Big{]} \end{eqnarray}
where capital Greek letters denote ten-dimensional indices, $\Phi$
is the dilaton, $H_{MNP}$ denotes the field strength of the
antisymmetric tensor field, which we shall ignore in the present
work, and $T$ is a tachyon field of mass $m^2 <0$. In our analysis
we have ignored higher than quadratic order terms 
in the tachyon potential. The
quantity ${\cal F}_{MNP\Sigma T}$ denotes the appropriate
five-form of type-$0$ string theory, with non trivial flux, 
which couples to the tachyon
field in the Ramond-Ramond (RR) sector via the function $f(T)=1 +
T + \frac{1}{2}T^2$.

{}From (\ref{actiontype0}) one sees easily the important r\^ole of
the five-form ${\cal F}$ in stabilizing the ground state. Due to
its special coupling with the quadratic $T^2$ term in Ramond-Ramond (RR)
sector of
the theory, it yields an effective mass term for the tachyon which
is positive, despite the originally negative $m^2$
contribution~\cite{type0}.
As mentioned previously, such a stability has recently been 
questioned in the context of string loop corrections,
but as we have mentioned previously this is rather a desirable
feature of the approach, in view of the claimed
cosmological instabilities.

As argued in \cite{dgmpp} 
{\it fluctuations of the brane worlds} involved in 
the construction of type-0 string theory result in 
{\it supercriticality} of the underlying $\sigma$-model,
with inevitable consequence 
the addition of the following 
term to the action
(\ref{actiontype0}) \ba \label{qterm}
 - \int d^{10} x
\sqrt{-G}e^{-2\Phi} Q(t)^2 
\ea 
where $\Phi$ is the dilaton field, and $Q(t)$ is the 
central-charge deficit of the non-equilibrium non conformal 
$\sigma$-model theory. The time here is identified with the 
(world-sheet zero mode of) the Liouville field, 
and the $t$ dependence of the central charge deficit is in accordance
with the concept of a Zamolodchikov C-function~\cite{zam}, 
a ``running central charge'' of a 
non-conformal theory, interpolating between two conformal 
(fixed point) theories. 

The sign of $Q(t)^2$  is positive if one assumes
supercriticality of the string~\cite{aben,ddk,grace,dgmpp}, 
which is the case of the model of \cite{dgmpp}. 
It is important to remark that 
in general, $Q^2(t)$ depends on the $\sigma$-model
backgrounds fields, being the analogue of Zamolodchikov's
$C$-function~\cite{zam}. 
As explained in \cite{dgmpp}, the explicit 
time dependence of $Q(t) $ reflects the existence 
of relevant operators in the problem,
other than the background fields considered in (\ref{actiontype0})
which are treated collectively in the present context.
Such operators have been argued to represent {\it initial} 
quantum fluctuations of the brane world. A plausible scenario, for instance, 
would be that the initial disturbance that takes the system out of equilibrium
is due to an impulse on the D3 brane worlds coming from either
a scattering off it of a macroscopic number of closed string bulk states
or another brane in scenaria where the bulk space is uncompactified 
(e.g. ekpyrotic universes {\it etc.}~\cite{ekpyrotic}).
For times long after the event, memory of the details of this process is
kept in the temporal evolution of $Q^2(t)$, which is 
determined self-consistently by means of the Liouville 
equations, as we shall see below.

We note in passing that such time-,
 and background field, -dependent deficits
also appear~\cite{schmid} if one views a standard 
critical string on a $d+1$-dimensional 
cosmological string background as a 
{\it non critical} $\sigma$-model propagating on  
a spatially-dependent 
$d$-dimensional background. 
In such a case the time dependence of the spatial coordinates
$x^i(t),~i=1, \dots d$, is attributed to the fact that 
the latter represent trajectories in the $d+1$-dimensional
space time. In this respect, the starting 
point is a $d$-dimensional non-conformal 
$\sigma$-model with only {\it spatially dependent} backgrounds. 
The extra time 
variable then is viewed as a time-like Liouville field 
which restores conformal invariance of the $d$-dimensional 
stringy $\sigma$-model. In this way, the model  
has a time-dependent (`running')
central charge 
deficit, being given essentially by the $d$-dimensional 
target-space effective action.
In our approach~\cite{emn,grace,dgmpp}, however, 
as we 
have stressed repeatedly, 
the dimensionality of the target space time is not increased
by the Liouville-dressing procedure.
Instead, the time (=Liouville) field itself readjusts its configuration
in a non-linear way in order to restore the conformal invariance 
broken by the fluctuations of the brane worlds in the type-0
string theory~\cite{emn,dgmpp}.

The ten-dimensional metric configuration we considered in \cite{dgmpp} 
was: 
\begin{equation}
G_{MN}=\left(\begin{array}{ccc}g^{(4)}_{\mu\nu} \qquad 0 \qquad 0 \\
0 \qquad e^{2\sigma_1} \qquad 0 \\ 0 \qquad 0 \qquad
e^{2\sigma_2} I_{5\times 5} \end{array}\right)
\label{metriccomp}
\end{equation}
where lower-case Greek indices are four-dimensional space time
indices, and $I_{5\times 5}$ denotes the $5\times 5$ unit matrix.
We have chosen two different scales for internal space. The field
$\sigma_{1}$ sets the scale of the fifth dimension, while
$\sigma_{2}$ parametrize a flat five dimensional space. In the
context of cosmological models, we are dealing with here, the
fields $g_{\mu\nu}^{(4)}$, $\sigma_{i},~i=1,2$ are assumed to
depend on the time $t$ only.

As we demonstrated in \cite{dgmpp}, a consistent background choice
for the flux form field will be that in which the flux is parallel to 
to the fifth dimension $\sigma_2$. This implies actually 
that the internal space is crystallized (stabilized) 
in such a way that this dimension is much larger than the 
remaining four $\sigma_1$. 

Upon considering the fields to be time dependent only, 
i.e. considering spherically-symmetric homogeneous backgrounds, 
restricting
ourselves to the compactification (\ref{metriccomp}), and assuming
a Robertson-Walker form of the four-dimensional metric, with scale
factor $a(t)$, the generalized conformal invariance conditions
and the Curci-Pafutti $\sigma$-model renormalizability constraint~\cite{curci}
imply a set of differential equations, which we solved numerically 
in \cite{dgmpp}.

The generic form of these equations reads~\cite{ddk,emn,dgmpp}:
\begin{equation} 
  {\ddot g}^i + Q(t){\dot g}^i = -{\tilde \beta}^i 
\label{liouvilleeq}
\end{equation} 
where ${\tilde \beta}^i$ are the Weyl anomaly coefficient of the 
stringy $\sigma$-model on the background $\{ g^i \}$. 
In the model of \cite{dgmpp} the set of $\{ g^i \}$ contains
graviton, dilaton, tachyon, flux and moduli fields $\sigma_{1,2}$ 
whose vacuum expectation values control the size of the extra dimensions.
The equations, then, have the following explicit form: 
\ba &~& -3\frac{\ddot a}{a} + {\ddot
\sigma}_1 + 5 {\ddot \sigma}_2 - 2 {\ddot \Phi} + {\dot
\sigma}_1^2  + 5 {\dot \sigma}_2^2 + \frac{1}{4}{\dot T}^2 +
e^{-2\sigma _1 + 2\Phi}f_5^2 f(T)=0~, \nonumber \\ &~&  {\ddot a}a
+ a{\dot a}\left(2Q + {\dot \sigma}_1 + 5{\dot \sigma}_2 - 2{\dot
\Phi}\right)+ e^{-2\sigma_1 + 2\Phi}f_5^2 f(T)a^2 =0~, \nonumber
\\ &~& {\ddot \sigma}_1 + 5{\dot \sigma}_1^2 + 3\frac{{\dot
a}}{a}{\dot \sigma}_1 + 2Q{\dot \sigma}_1 + 5{\dot \sigma}_1{\dot
\sigma}_2 -  2{\dot \sigma}_1{\dot \Phi}+ e^{-2\sigma_1 +
2\Phi}f_5^2~f(T)=0 ~, \nonumber \\ &~&  3{\ddot \sigma}_2 + 9{\dot
\sigma}_2^2 + 3\frac{{\dot a}}{a}{\dot \sigma}_2 + 2Q{\dot
\sigma}_2 + {\dot \sigma}_1{\dot \sigma}_2 -  2{\dot
\sigma}_2{\dot \Phi}- e^{-2\sigma_1 + 2\Phi}f_5^2~f(T)=0~,
\nonumber
\\ &~&  2{\ddot T} + 3\frac{{\dot a}}{a}{\dot T} + Q~{\dot T} +
{\dot \sigma}_1 {\dot T} + 5{\dot \sigma}_2 {\dot T} - 2{\dot
T}{\dot \Phi} + \nonumber \\ &~& m^2T - 4 e^{-2\sigma_1 +
2\Phi}f_5^2 f'(T)=0~, \nonumber \\ &~&  {\ddot \Phi} + Q{\dot \Phi
} + 6\frac{{\dot a}}{a} + 6\frac{{\dot a}^2}{a^2} + \nonumber
\\ &~&2\left[-{\ddot \sigma}_1 - \frac{3{\dot a}}{a}{\dot \sigma}_1
-5{\ddot \sigma}_2 - 15\frac{{\dot a}}{a}{\dot \sigma}_2 - {\dot
\sigma}_1^2 - 15 {\dot \sigma}_2^2 - 5 {\dot \sigma}_1{\dot
\sigma}_2 - \right.\nonumber \\ &~&2\left.{\dot \Phi}^2 + 2{\ddot
\Phi} + 6\frac{{\dot a}}{a}{\dot \Phi} + 2{\dot \sigma}_1{\dot
\Phi} + 10{\dot \sigma}_2{\dot \Phi} \right] - \frac{1}{4}{\dot
T}^2 + \frac{1}{4}m^2T^2 + Q^2=0~,\nonumber \\ &~&
C_{5}=e^{-\sigma_1 + 5 \sigma_2}f(T)f_5~, \nonumber \\ &~&
{\Phi}^{(3)} + Q{\ddot \Phi} + {\dot Q}{\dot \Phi} + 12\frac{{\dot
a}}{a^3}\left(a{\ddot a} + {\dot a}^2 + Q~a{\dot a}\right) -{\dot
T}({\ddot T}+ Q{\dot T}) + \nonumber \\ &~& 4{\dot
\sigma}_1({\ddot \sigma}_1 + 2{\dot \sigma}_1^2 + Q{\dot
\sigma}_1) + 20{\dot \sigma}_2({\ddot \sigma}_2 + 2{\dot
\sigma}_2^2 + Q{\dot \sigma}_2)=0 \label{eqsmotion} \ea where
$f'(T)$ denotes functional differentiation with respect to the
field $T$, the overdot denotes time derivative, 
with respect to the $\sigma$-model frame, 
and $\Phi^{(3)}$ denotes
triple time derivative.

As argued in \cite{dgmpp} such equations 
correspond to solutions of equations of motion 
derived from a
ten-dimensional effective action. This is an important
and non-trivial consequence of the gradient flow property of the 
$\sigma$-model ${\tilde \beta}^i$ functions, according to which:
\begin{equation} 
    {\tilde \beta}^i = {\cal G}^{ij}\frac{\delta {\cal F}[g]}{\delta g^j}
\label{flow}
\end{equation}
where the flow functional  ${\cal F}[g]$ is essentially 
the target-space effective 
action, depending on the background configuration under consideration. 

An equivalent set of equations (in fact at
most linear combinations) come out from the corresponding
four-dimensional action after dimensional reduction. Of course
this reduction leads to the string 
or $\sigma$-model
frame, in which there are dilaton exponential 
factors in front of the Einstein term in the action. 
We may turn to the Einstein frame,
in which such factors are absent, and the 
Einstein term is canonically normalized,  
through the
 transformation
 \begin{equation}
 g_E=e^{-2\Phi+\sigma _1 + 5\sigma _2}g
 \end{equation}
 In this frame the line element is
 \begin{equation}
 ds_E^2 =  -e^{-2\Phi+\sigma _1 + 5\sigma _2}dt^2
 +a^2(t) e^{-2\Phi+\sigma _1 + 5\sigma _2}(dr^2 +r^{2}d\Omega^{2})
 \end{equation}
Therefore to discuss cosmological evolution we
 have to pass to the cosmological time defined by
\begin{equation}\label{l1}
dt_E =  e^{-\Phi+\frac{ \sigma _1 + 5 \sigma _2}{2}}dt
\end{equation}\
Then the line element, for a spatially flat universe, which we assume here
motivated by the CMB data~\cite{evidenceflat},  
becomes: 
\begin{equation}\label{l2}
  ds_E^2 = -dt_E^2+a_E^2(t_E)(dr^2+r^{2}d\Omega^{2})
\end{equation}
  with 
\begin{equation}
a_E(t_E)=
  e^{-\Phi+\frac{\sigma _1 + 5\sigma _2}{2}}a(t(t_E))~.
\label{l3}
\end{equation} 
Recalling 
the notation~\cite{dgmpp}: 
\begin{equation} 
a=e^{b(t) t}
\end{equation} 
in the $\sigma$-model frame, 
one has the following useful relations between the Einstein 
and $\sigma$-model frames, to be used here: 
\begin{eqnarray}
&~& \frac{da_E}{dt_E} = e^{b}\left(-{\dot \Phi} + \frac{{\dot \sigma}_1 + 
5{\dot \sigma}_2}{2} + {\dot b}\right)~, \nonumber \\
&~& \frac{d^2a_E}{dt_E^2} = e^{b}e^{\Phi - \frac{\sigma_1 + 5\sigma_2}{2}}
\left[{\dot b} \left (-{\dot \Phi}+ \frac{{\dot \sigma}_1 + 5{\dot \sigma}_2}{2} + {\dot b}\right) -{\ddot \Phi} + \frac{{\ddot \sigma}_1 + 5{\ddot \sigma}_2}{2} + {\ddot b}\right]~.
\label{einstsigma}
\end{eqnarray} 
The Hubble parameter reads: 
\begin{equation} 
H(t_E)  \equiv \frac{\frac{d{a}_E}{dt_E}}{{a}_E}
=e^{\Phi - \frac{\sigma_1 + 5\sigma_2}{2}}\left(-{\dot \Phi}
+ \frac{{\dot \sigma}_1 + 5{\dot \sigma}_2}{2} + {\dot b}\right)
\label{hubbleparameter}
\end{equation} 
while the deceleration parameter of the Universe (\ref{decel}) 
acquires the form: 
\begin{equation} 
q=-\frac{{\dot b}\left(-{\dot \Phi} + 
\frac{{\dot \sigma}_1 + 5{\dot \sigma}_2}{2} + {\dot b}\right)
-{\ddot \Phi} + \frac{{\ddot \sigma}_1 + 5{\ddot \sigma}_2}{2} + {\ddot b}}
{\left(-{\dot \Phi} + 
\frac{{\dot \sigma}_1 + 5{\dot \sigma}_2}{2} + {\dot b}\right)^2}
\label{decel2b} 
\end{equation} 
Finally the Einstein-frame 
``vacuum'' energy is related to the central charge deficit~\cite{dgmpp}
\begin{equation} 
\Lambda_E = e^{2\Phi - \sigma_1 - 5\sigma_2}Q^2(t) 
\label{vacuumenergy} 
\end{equation} 
As discussed in \cite{dgmpp}, due to the non-equilibrium nature of the 
non-critical string Universe, which has not yet relaxed to its 
ground state, $\Lambda_E$ 
should be considered rather as an effective potential,
in much the same way as the potential of a (non-equilibrium)
quintessence field, whose r\^ole is played here the 
dilaton $\Phi$~\cite{emnsmatrix,dgmpp}. 

We should also remark that we have adopted the usual Kaluza-Klein
reduction without considering the extra dimensions compactified.
Nevertheless the equations we are interested in and the results we
will discuss in the following section are not modified if we had
considered a compact six-dimensional space instead. In that case
$e^{2\sigma_1}$ and $e^{2\sigma_2}$ should correspond to radii of
the compact space. Note also that in the string frame we have the
exponential $e^{-2\Phi+\sigma_1+5\sigma_2}$ instead of
$e^{-2\Phi}$, since we allow time dependence of the volume of the
compact space.

The numerical solution we have found 
is supported by analytical considerations
for the asymptotic field modes (late cosmological-frame times $t \to \infty$).
We followed an iterative method
of solving the system of these equations. 
The starting point of the iteration procedure is the solution of
 the linear system with the correct asymptotic behaviour.
Then we insert the linear solution 
into the system of equations and keep only the
linear part to get the general solution.
For details and results we refer the reader in \cite{dgmpp}.

%%%%%%%%%%%%%%%%%%%%%%%%%%%%%%%%%%%%%%%%%%%%%%
\begin{figure}[h]
\centering
\includegraphics[scale=0.5]{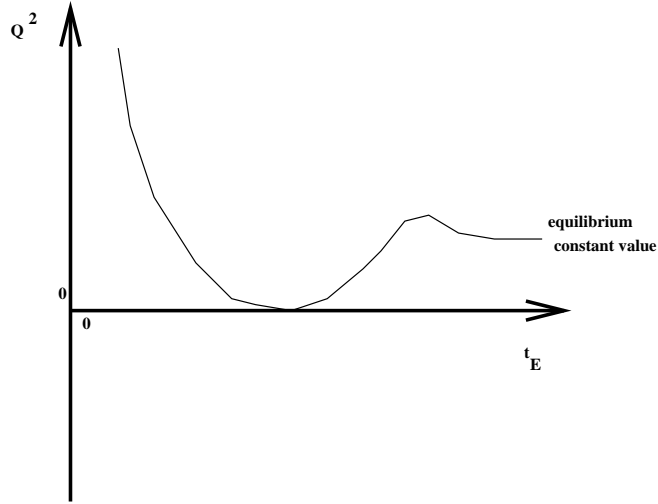}
%%\centerline{\hbox{\psfig{figure1b=figure1b.eps,height=10cm}}}
\caption {The evolution of the central charge deficit $Q^2$
in the Einstein frame. Immediately after inflation, 
the deficit passes through a phase where it 
first vanishes, and then oscillates before 
relaxing to an equilibrium constant value asymptotically.} 
\label{centralcharge:fig}
\end{figure}
%%%%%%%%%%%%%%%%%%%%%%%%%%%%%%%%%%%%%%%%%%%%%%

%%%%%%%%%%%%%%%%%%%%%%%%%%%%%%%%%%%%%%%%%%%%%%
\begin{figure}[h]
\centering
\includegraphics[scale=0.5]{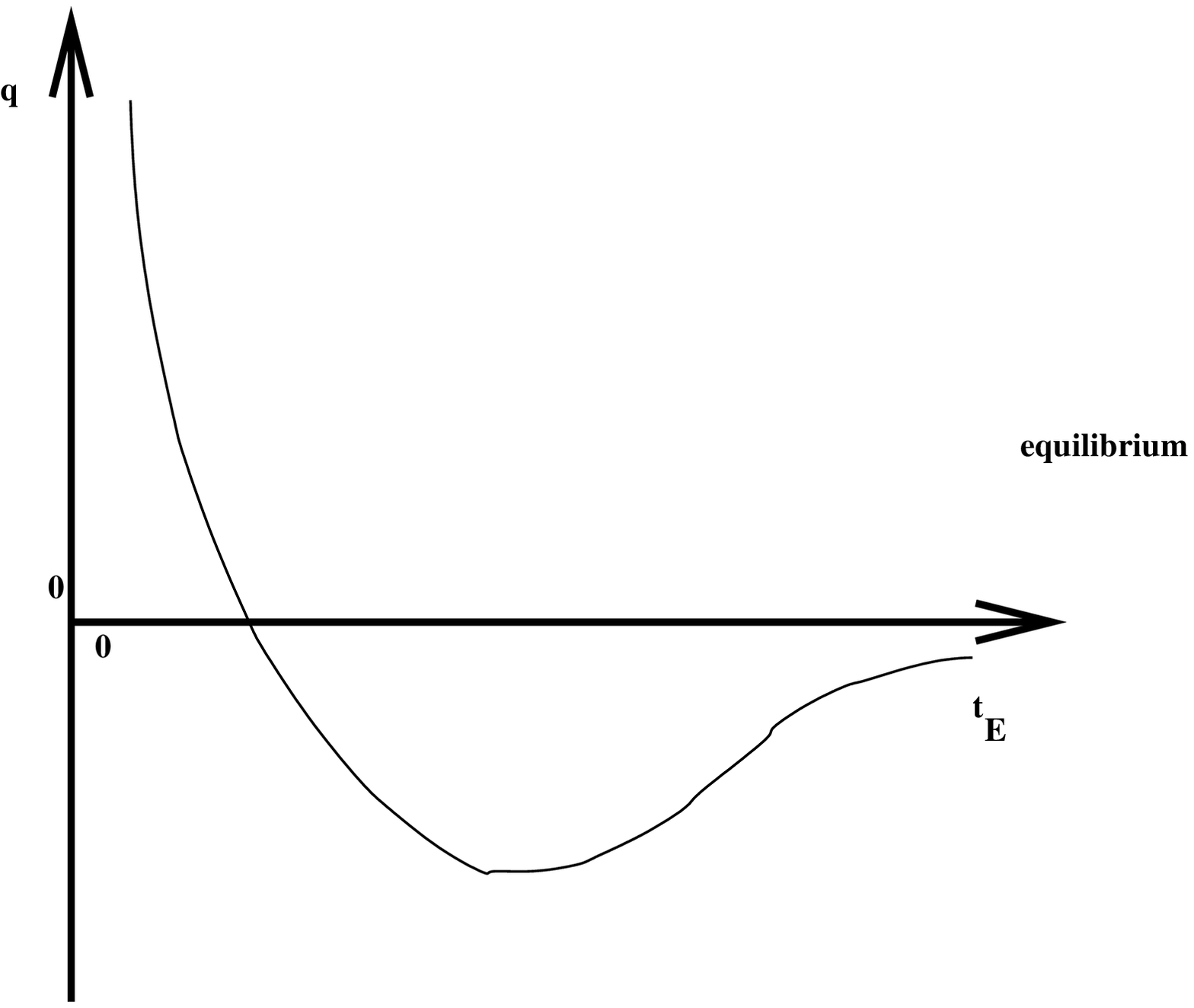}
%%\centerline{\hbox{\psfig{figure1b=figure1b.eps,height=10cm}}}
\caption {The evolution of the decelerating parameter $q$  
of the type-0 string Universe in the Einstein frame.}
\label{decel:fig}
\end{figure}
%%%%%%%%%%%%%%%%%%%%%%%%%%%%%%%%%%%%%%%%%%%%%%

The most important physical results of our analysis 
may be summarized as follows: 
our solution 
demonstrates  
that the scale factor of the Universe, 
after the initial singularity, enters a
short inflationary phase and then, in a smooth way, goes into 
a flat
Minkowski spacetime with a linear dilaton for 
asymptotically long $\sigma$-model times $t \rightarrow \infty$
in the $\sigma$-model frame.
Equivalently, in the Einstein frame, this asymptotic  
string theory corresponds to a string propagating into a 
a linearly expanding, non-accelerating Universe, 
with a (negative valued) dilaton that varies logarithmically 
with the Robertson-Walker (Einstein frame) time. 
Note that this is a consistent $\sigma$-model background~\cite{aben}  
in the sense of satisfying modular invariance and factorization
of $S$-matrix elements. This is actually one of the most important 
points of our work in \cite{dgmpp}, namely that 
there is a smooth exit from the de Sitter phase in such a way
that one can {\it appropriately define} asymptotic on-shell 
states, and hence an $S$-matrix.  
The 
fields $\sigma_{1}$ and $\sigma_{2}$ which
parametrize the internal space have an interesting behaviour. The
field $\sigma_{1}$, which sets the scale of the fifth dimension,
during inflation contracts until it reaches a constant value.
After inflation, it maintains this value, until the universe
evolves to the above mentioned asymptotically flat spacetime 
in the $\sigma$-model frame. 
The field $\sigma_{2}$
which parametrizes the conformally flat five-dimensional space freezes to a
constant value which is much smaller than that of 
the fifth dimension. Thus we see that, in our model,  
a cosmological
evolution may lead to different scales for the extra
dimensions. 
It is important to notice, that this
difference in scales of the extra dimensions is due to the fact
that in our theory the gravity is very weak asymptotically.
A phenomenologically important feature of our model is that
the vacuum energy, determined by the central-charge deficit,
relaxes to zero asymptotically in a way which is reminiscent
of quintessence models, with the r\^ole of the 
quintessence field played by the dilaton.
During the {\it relaxation process} the string theory remains
consistent as a conformal $\sigma$-model, since the 
target time, which here plays the r\^ole of the Liouville 
field, restores the broken conformal invariance~\cite{emn,grace,dgmpp}.

The Cosmology of the model in the late time phase,
where the extra dimensions have been stabilized to their
equilibrium values and the tachyonic mode has relaxed to a constant value
(assumed zero by normalization)~\cite{dgmpp}, may be summarized
diagrammatically in figures \ref{centralcharge:fig}
and \ref{decel:fig}. In these figures  
we give the evolution (in the Einstein frame) of the
central charge deficit and the 
deceleration parameter with time.
The figures refer to the numerical solution of \cite{dgmpp}.

From figure \ref{centralcharge:fig} we observe 
that, 
immediately after inflation, the central charge deficit $Q^2$ 
passes through a
{\it metastable point} where it vanishes. 
This has to do with the fact 
that at this point 
the square root $Q$ changes sign~\footnote{This change of sign of $Q$ is 
irrelevant for Liouville $\sigma$-model physics, although important
for target space physics as we shall discuss later on.
Indeed, in the original Liouville
$\sigma$-model
Lagrangian one encounters only $Q^2$ factors~\cite{ddk}. To obtain 
a canonical normalization for the Liouville $\sigma$-model
kinetic term one needs only to perform the redefinition~\cite{ddk} 
$\phi \to |Q(t)|\phi$ to the Liouville mode, which is thus unaffected by a 
sign 
change of $Q(t)$, with $t$ the world-sheet 
zero mode of the Liouville field $\phi$.}, as
can also be confirmed analytically from the 
corresponding expression for the linearized solution of \cite{dgmpp}. 
After this point the central charge deficit 
oscillates (as a function of time) before it relaxes to its constant asymptotic
value, which should be one of the values for which the conformal theory of 
\cite{aben} is valid. The oscillatory nature is consistent with the 
time-like signature of the Liouville mode as explained in \cite{dgmpp}. 
This behaviour of the central charge 
deficit is important in determining the 
evolution of the energy densities of the type-$0$ 
string Universe, as we shall discuss 
later on.

We also observe from fig. \ref{decel:fig} that, 
after its graceful exit from the inflationary phase, 
the type-$0$ non-critical string Universe 
passes first through a decelerating phase, 
which is then succeeded by an accelerating
one, 
before the Universe relaxes asymptotically to is steady state (equilibrium)
value. 

It is the purpose of this article to discuss the precise phenomenology
of this Universe, and compare it with current 
observational evidence~\cite{evidomegal}. To this end 
we shall computing the value of the deceleration parameter 
at the ``present era''. The concept of `present era' 
will be defined appropriately in section 3.

Before doing so, we would like to make 
an important comment concerning the initial 
singularity which characterizes our solution. 
The singularity is a general feature of the equations of
the form (\ref{eqsmotion}). The singularity
may be a true one, in case one thinks of the initial 
brane fluctuation which caused the non-criticality as a 
{\it catastrophic event} (e.g. a situation in the ekpyrotic
universe scenario~\cite{ekpyrotic}), or it may be removable,
in case the initial fluctuation is a quantum fluctuation in the context of 
the M-theory. In the latter case, 
removing the singularity is probably a matter of
a full quantum description of the theory,
which at present is not available. We note 
at this stage, however, the possibility
of deriving smooth cosmological solutions in string theory,
without initial singularities, by including in the action
higher curvature terms (e.g. quadratic 
of Gauss-Bonnet type~\cite{art}), which are part of the quantum
corrections, in the sense of being generated by 
including string-loop corrections in the effective action.
For our purposes here, such issues are not directly relevant.

\section{Present Era Phenomenology of the non-critical type-0 
String Universe: Compatibility with supernovae Ia observations} 

Having obtained numerical evidence (c.f. fig. \ref{decel:fig}) 
from the full numerical solution
on the existence of an accelerating phase of the type-$0$ string Universe, 
before its 
final relaxation into the critical string regime, 
we now proceed to study {\it analytically} 
this phase. 
In particular, in this section, we shall be interested 
in studying the {\it precise} behaviour, as well as 
estimating the {\it order of magnitude}, of the 
{\it deceleration parameter q}
(\ref{decel2b})
for Einstein frame times long after the initial 
fluctuation of the $D3$ brane worlds. This would allow
comparison with the current observations. 

In this late time regime  
the 
various fields can be approximated well by their {\it linearized solution},
described in detail in \cite{dgmpp}. 
Recalling that for the Einstein-frame time one has
\begin{equation}\label{einsteintime} 
t_E = t_{E,0} + \int_{t_0}^t d\tau \left(e^{-\Phi(\tau)+ \frac{\sigma_1(\tau)
+ 5\sigma_2(\tau)}{2}}\right)
\end{equation}
we mention that,
for the numerical solution of \cite{dgmpp},   
the various field modes can be approximated sufficiently well by the
linear solution for $t_0 \ge 25$ in string units of time. 

We will also assume that at present times, long after the inflationary period,
which we are interested
in here, 
the extra dimensions,
as well as the tachyon field, are already frozen to their constant
asymptotic values $\sigma_{1,0} \equiv s_{01}$, $\sigma_{2,0}\equiv s_{02}$.
The validity of this assumption will be discussed later on.
For the tachyon field, this freezing-out behaviour implies that 
$f(T)=1+T+T^2/2 \rightarrow 1$, while the freezing out of the extra dimensions 
implies that 
the ``volume" $V_6 \equiv e^{s_{01} + 5s_{02}}$ 
of the extra dimensions can be absorbed, as usual
after compactifcation, 
in the four-dimensional 
gravitational coupling constant.
This will be assumed in what follows, which means that, from now on, 
the string and the Einstein times in (\ref{einsteintime}) 
will be related only through the dilaton
field, without explicit involvement of $V_6$.
The freezing out of the extra dimensions and the tachyon 
to their constant (equilibrium) values 
implies that any non-trivial 
dependence on the fields $\sigma_{1}, \sigma_{2}$ and $T$ 
disappears from the equations satisfied by the other
fields. 

Keeping, therefore, only the modes $a,~Q,~\Phi~,f_5$ in 
the remaining equations (\ref{eqsmotion}), 
we observe that they acquire the form:
\begin{eqnarray} 
&~& \frac{C_5^2 e^{2\sigma_{01}}}{2V_6^2}e^{2\Phi} -3\frac{\ddot a}{a}-2{\ddot \Phi}=0~,
\nonumber \\
&~& \frac{C_5^2 e^{2\sigma_{01}}}{2V_6^2}e^{2\Phi} + (2Q-2{\dot \Phi})\frac{\dot a}{a}
+ \frac{\ddot a}{a}=0~, \nonumber \\
&~& \Phi^{(3)} + Q{\ddot \Phi}+ {\dot Q}{\dot \Phi}=0~\qquad ({\rm Curci-Paffuti~relation})~, \nonumber \\
&~& f_5 =\frac{C_5e^{2s_{01}}}{V_6}~\qquad ({\rm five-form
~equation})~,\nonumber \\
&~& -Q^2 -Q{\dot \Phi} + 4{\dot \Phi}^2 -6\frac{\ddot a}{a} -
12\frac{\dot a}{a}{\dot \Phi}-5{\ddot \Phi} =0~ ({\rm dilaton~equation}).
\label{eqmotion2}
\end{eqnarray} 
Note that the five-form field is also frozen to a constant value.
In the Curci-Paffuti relation we have dropped out the combination
$12 \frac{{\dot a}}{a^3}(a {\ddot a}+{\dot a}^2+Qa{\dot a})$, which
is very small compared to the other terms, for the time period
we are interested in.

The Curci-Paffuti relation in (\ref{eqmotion2}) can be integrated for
the field $Q$ and the solution respecting the asymptotic behaviour
$Q \rightarrow q_0$ is:
\begin{equation}
Q=-\frac{F_1q_0}{{\dot \Phi}}-\frac{{\ddot \Phi}}{{\dot \Phi}}
\label{eqforq}
\end{equation}
where $F_{1}$ is a positive constant and 
${\dot\Phi} \rightarrow -F_1$ (asymptotically linear dilaton).
The system of the remaining three non-trivial equations then
reads:
\begin{eqnarray} 
&~&\alpha_1  + 0\times \frac{\dot a}{a} - 3\frac{\ddot a}{a} 
= 0~, \nonumber \\
&~& \beta_1 + (2Q-2{\dot \Phi})\frac{\dot a}{a} + \frac{\ddot a}{a} =0~, \nonumber \\
&~& \gamma_1 -12{\dot \Phi}\frac{\dot a}{a} - 6\frac{\ddot a}{a} = 0~, \nonumber \\
&~& {\rm where}~\alpha_1 \equiv \frac{C_5^2}{2V_6^2e^{-2s_{01}}}e^{2\Phi}-
2{\ddot \Phi}, \quad 
\beta_1 \equiv  \frac{C_5^2}{2V_6^2e^{-2s_{01}}}e^{2\Phi}, \nonumber \\
&~& \gamma_1 \equiv -Q^2 -Q{\dot \Phi}+4{\dot \Phi}^2 -5{\ddot \Phi}
\end{eqnarray}
Compatibility requires the vanishing of the determinant
\begin{equation}
0 = {\rm det}\left(\begin{array}{ccc}\alpha_1 \qquad 0 \qquad -3 \\
\beta_1 \qquad 2Q-2\dot \Phi \qquad 1 \\\gamma_1 \qquad -12\dot \Phi \qquad -6 \end{array}\right)
\label{resolvent}\end{equation}
which yields an equation for $\Phi$, whose solution, by compatibility,
must yield the linear solution $\Phi \simeq f_0 -F_1 t$ for large times $t$.
Indeed, this is what happens, as we shall demonstrate below. 

We first observe that to linear order in the fields
one obtains:
\begin{equation} 
Q = q_0 + \frac{q_0}{F_1}(F_1 + \dot \Phi).
\label{ccd}
\end{equation} 
{}From this we, therefore, see that $Q \to q_0$ 
if and only if ${\dot \Phi} \rightarrow 
-F_1$, which, as  we shall see below,
is true.

Indeed, 
from the resolvent (\ref{resolvent}), keeping only 
terms linear in the fields, we obtain for $\Phi(t)$:
\begin{equation}\label{res2}
{\ddot \Phi}(t) + \alpha^2 A^2 e^{2\Phi(t)} + \beta^2 F_1 (F_1 + {\dot \Phi})=0
\end{equation}
where $\alpha^2 = \frac{11 + \sqrt{17}}{2(3 + \sqrt{17})}$, 
$\beta^2=\frac{17 + 5\sqrt{17}}{3 + \sqrt{17}}$ are numerical constants of 
order ${\cal O}(1)$, and $A^2 \equiv \frac{C_5^2e^{2s_{01}}}{2V_6^2}$.

For times long after the initial fluctuations, 
such as the present times, where the linear approximation is valid, 
the term 
$\beta^2 F_1 (F_1 + {\dot \Phi})$ is hierarchically small,  
and may
be neglected in the dilaton equation (\ref{res2}), yielding 
$\alpha^2 A^2e^{2\Phi}+ {\ddot \Phi} \simeq 0$. This is solved for:
\begin{equation} \label{dilaton} 
\Phi (t) =-{\rm ln}\left[\frac{\alpha A}{F_1}{\rm cosh}(F_1t)\right],
\end{equation}
with $F_1$ a positive constant.
For large times $F_1 t \gg 1$ (in string units) 
one therefore recovers the
linear solution for the dilaton, thereby demonstrating the self-consistency
of the approach: 
$\Phi \sim f_0 -F_1 t$, $F_1 =|f_1| >0$~\footnote{From (\ref{dilaton}) 
we thus observe that the asymptotic weakness of 
gravity in this Universe~\cite{dgmpp} is due to the smallness of 
the internal space $V_6$ as compared with the flux $C_5$ of the 
five form field, $f_0 \sim {\rm ln}V_6$ (c.f. previous expression for $A$).
We shall come back to this important point later on.}.

Defining the Einstein frame time $t_E$ as
\begin{equation}\label{sigmatime} 
t_E=\int ^t e^{-\Phi (z)}dz 
\end{equation}
we get 
\begin{equation}
t_E=\frac{\alpha A}{F_1^2}sinh(F_1 t).
\end{equation}
The string frame time $t$, can be expressed in terms of $t_E$ as:
\begin{equation}
t=\frac{1}{F_1} ln \left[ \sqrt{1+\frac{F_1^4}{\alpha ^2 A^2} t_E^2}
+ \frac{F_1^2}{\alpha A} t_E \right].
\end{equation}
In terms of the Einstein-frame time (\ref{sigmatime}) 
one obtains a logarithmic time-dependence
~\cite{aben} for the dilaton
\begin{equation} 
\Phi _E = {\rm const} -{\rm ln}t_E~,
\label{einsteindil} 
\end{equation}

For this behaviour of $\Phi$, the central charge deficit 
(\ref{ccd}) tends to a constant value $q_0$.
This value must be, for consistency of the underlying string theory,
{\it one of the discrete values} obtained in \cite{aben}, 
for which the factorization property (unitarity) of the 
string scattering amplitudes 
occurs. Notice that this asymptotic 
string theory, 
with 
a constant (time independent) central-charge deficit,  
$Q^2 \propto c-25 $ (or $c-9$ for superstring) 
is considered an {\it equilibrium} situation,
where an $S$-matrix can be defined for specific (discrete)
values of the central charge $c$. The standard
critical (super)string corresponds to central charge 
$c =25$ (=9 for superstrings)~\cite{ddk,aben}.

We would like at this point to 
go back for a moment and justify  
the validity of the assumption
that the extra dimensions are frozen, and hence any non-trivial
dependence on the fields $\sigma_{1,2}$ can be ignored.
The evolution of the extra dimensions
is determined by the third and forth 
of the equations (\ref{eqsmotion}) 
\begin{eqnarray} 
&~& A^2 e^{2\Phi} -2 {\dot \sigma}_1 \left(
 {\dot \Phi}   + \frac{{\ddot \Phi}}{{\dot \Phi}} + 
\frac{q_0 F_1}{{\dot \Phi}}\right) + {\dot \sigma}_1 {\dot \sigma}_2
+5{\dot \sigma}_1^2 +  {\ddot \sigma}_1 = 0 \nonumber \\
&~&  -A^2 e^{2\Phi} -2 {\dot \sigma}_2 \left(
 {\dot \Phi}   + \frac{{\ddot \Phi}}{{\dot \Phi}} + 
\frac{q_0 F_1}{{\dot \Phi}}\right) + {\dot \sigma}_1 {\dot \sigma}_2
+9{\dot \sigma}_2^2 + 3 {\ddot \sigma}_2 = 0
\label{evolextradim1}
\end{eqnarray} 

 In the range where the linear approximation is valid this system 
 takes the form

\begin{eqnarray} 
&~& {\ddot \sigma}_1 + 2(q_0 + F_1){\dot \sigma}_1 + 
\frac{F_1^2}{\alpha ^2 cosh^2(F_1t)}=0~, \nonumber \\
&~& 3{\ddot \sigma}_2 + 2(q_0 + F_1){\dot \sigma}_2 - 
\frac{F_1^2}{\alpha ^2 cosh^2(F_1t)}=0~.
\label{linearnew}
\end{eqnarray}

 The above system has indeed regular solutions of the type
 assumed above, namely solutions that freeze out to constant  values
relatively quickly after inflation. 
To see this,
let us concentrate for definiteness to the first of the 
equations (\ref{linearnew}).
The second equation can be analysed in a formally identical way. 

Integrating the first of equations (\ref{linearnew}) 
we obtain
$$ {\dot \sigma}_1 + 2(q_0 + F_1)\sigma_1 + \frac{F_1}{\alpha^2}tanh(F_1 t)=
c_1, \qquad c_1={\rm integration~constant}~.$$
Weighting this equation by $e^{2(q_0 + F_1)t}$ and integrating once more 
we obtain: 
\begin{equation} 
\sigma_1 +\frac{1}{\alpha^2}{\rm ln}cosh(F_1 t)-\frac{1}{\alpha^2}e^{-2(q_0 + F_1)t}\int dt' e^{2(q_0 + F_1)t'}{\rm ln}cosh(F_1 t') 
= \frac{c_1}{2(q_0 + F_1)} + e^{-2(q_0 + F_1)t}c_2~, 
\end{equation}
where $c_{1,2}$ are integration constants. We shall be interested in the regime
of long times $F_1 t \gg 1$ (in string units) after inflation. 
In this regime the $t$-integration  
can be done by approximating ${\rm ln}cosh(F_1 t) \simeq F_1 t$.
After some straightforward algebra we then obtain:
\begin{equation} 
\sigma _ 1 \to s_{01} = \frac{c_1/F_1 - 1}{2(1 + q_0/F_1)\alpha^2} + 
{\cal O}\left(e^{-2(q_0 + F_1)t}c_2\right)
\end{equation} 
where $q_0, F_1 > 0$. 
The reader should bear in mind that the solution of \cite{dgmpp} requires  
\begin{equation} 
\left(\frac{q_0}{F_1}\right) = \frac{1}{2}(1 + \sqrt{17}) \simeq 2.56~. 
\label{ci} 
\end{equation} 
This relation stems from 
the dilaton equation, which guarantees that the Liouville-dressed
theory preserves conformal invariance, as discussed in detail in 
\cite{dgmpp}.  

In a similar way one finds that the field $\sigma_2$ asymptotes
quickly to 
\begin{equation}
\sigma_2 \to s_{02} =   \frac{c_1'/F_1 + 1 }{2(1 + q_0/F_1)\alpha^2} + 
{\cal O}\left(e^{-\frac{2}{3}(q_0 + F_1)t}c_2'\right)
\end{equation} 
where $c_{1,2}'$ are appropriate integration constants. 

The constants in the equilibrium values $s_{0i}~, i=1,2$ can be chosen 
in such a way~\cite{dgmpp} that the size of the 
$\sigma_1$ dimension is larger
than the rest. 
We also remind the reader at this point 
that in this regime of large string-frame times 
$F_1 t \gg 1$ the dilaton can be approximated by its linear form 
$\Phi \sim -F_1 t$. This completes our digression on the 
self-consistency of the approximation of ignoring the effects of the 
$\sigma_{i}$ fields for large times $F_1 t \gg 1$. 

We now come to discuss the behaviour of the scale factor $a(t)$. 
 From the second of (\ref{eqmotion2}) one obtains  
the evolution equation for the scale factor, 
\begin{equation} 
\frac{\ddot a}{a} + (2q_0 + F_1)\frac{\dot a}{a} + \frac{F_1^2}{\alpha^2}\frac{1}{{\rm cosh}^2(F_1t)} = 0~, 
\end{equation} 
This can be easily solved by means of the Gauss hypergeometric 
functions $~_2F_1$; in the 
Einstein frame one has:
\begin{eqnarray}\label{scalefactoreinst}  
&~& a_E(t_E) = \frac{F_1}{\gamma}(\sqrt{1 + \gamma^2 t_E^2})\left[
C_1 \left(\frac{\sqrt{1 + \gamma^2 t_E^2}}{\gamma t_E + \sqrt{1 + \gamma^2 t_E^2}}\right)
^{\frac{F_1 + q_0}{F_1}} \times \right. \nonumber \\
&~& \left. _2F_1\left(\frac{1}{4}(1-\frac{2(F_1 + q_0)}{F_1}
-\frac{\sqrt{4 + \alpha^2}}{\alpha})~, \right. \right. \nonumber \\
&~& \left. \left. \frac{-2\alpha(F_1 
+ q_0)+F_1(\alpha+\sqrt{4+\alpha^2})}{4F_1 \alpha}~,
 1-\frac{F_1 + q_0}{F_1}~, \frac{1}{1 + \gamma^2 t_E^2}\right) + 
\right. \nonumber \\
&~& \left. C_2 \left[\sqrt{1 + \gamma^2 t_E^2} + 
\gamma t_E \right]^{-\frac{F_1 + q_0}{F_1}}~_2F_1\left(\frac{1}{4}(1+
\frac{2(F_1 + q_0)}{F_1}-\frac{\sqrt{4 + \alpha^2}}{\alpha})~, \right. \right. 
\nonumber \\
&~& \left. \left. \frac{2\alpha(F_1 + q_0) + F_1(\alpha+\sqrt{4+\alpha^2})}
{4F_1\alpha}~, 1 + \frac{F_1 + q_0}{F_1}~, \frac{1}{1 + \gamma^2 t_E^2}\right)\right] 
\end{eqnarray} 
where $C_{1,2}$ are integration constants, 
and 
\begin{equation} 
\gamma \equiv \frac{F_1^2}{\alpha A}~, \qquad A \equiv \frac{|C_5|e^{s_{01}}}{V_6}=
|C_5|e^{-5s_{02}}
\label{defA}
\end{equation}
Notice the independence of $A$ on the large compact dimension $s_{01}$.
This will play an important physical r\^ole as we shall discuss later on.

For large $t_E$, e.g. present cosmological time values, 
one has
\begin{equation} 
a_E(t_E) \simeq \frac{F_1}{\gamma}\sqrt{1 + \gamma^2 t_E^2}
\end{equation}
For very large (future ) times $a(t_E)$ scales linearly
with the Einstein-frame cosmological time $t_E$~\cite{dgmpp},
and hence the cosmic horizon (\ref{cosmhorizon}) disappears,
thereby allowing the proper definition of asymptotic
states and thus a scattering matrix. 
Asymptotically therefore in time,  
the Universe relaxes to its ground-state equilibrium situation,
and the non-criticality of the string, caused by the initial fluctuation,
disappears, making room for a critical (equilibrium) string Universe.

The Hubble parameter (\ref{hubbleparameter}) reads for large $t_E$ 
\begin{equation}\label{hubble2} 
H(t_E) \simeq \frac{\gamma^2 t_E}{1 + \gamma^2 t_E^2}
\end{equation}
while the deceleration parameter (\ref{decel}) in the same regime 
of $t_E$ becomes:
\begin{equation} 
q(t_E) \simeq -\frac{1}{\gamma^2 t_E^2}
\label{decel4}
\end{equation}
Finally, the ``vacuum energy'' reads:
\begin{equation} 
\Lambda (t_E) \simeq \frac{q_0^2 \gamma^2}{F_1^2 ( 1 + \gamma^2 t_E^2)}
\label{cosmoconst}
\end{equation}

From (\ref{hubble2}), (\ref{cosmoconst}) we observe that 
one can match the present-era values quite straightforwardly, as expected
by naive dimensional analysis. On the other hand, the dimensionless
deceleration parameter $q$ (\ref{decel4}), although negative, 
appears to be 
extremely suppressed as compared to the order one value inferred
from the best fit of the supernova data (\ref{bestfit}),(\ref{decel2}).
We should remark, however, that, 
due to the current uncertainties in the data~\cite{sarkar}, 
this is not necessarily in contradiction with the current 
observations for a non-zero cosmological constant.
Nevertheless, if one takes the best fit Universe (\ref{bestfit}) literally,
then it seems that in order to match all the data one should 
obtain an order one  
deceleration parameter of the type 0 string Universe 
(c.f. discussion following (\ref{decel2})). 

In our case this is possible, since as we 
observe from (\ref{scalefactoreinst}), in the Einstein frame
the scale factor of our type-0 non-critical string Universe
is only a function of the combination $\gamma t_E$.
If, therefore, one {\rm defines} the  
{\it present era} by the time regime
\begin{equation}
\gamma \sim t_E^{-1} 
\label{condition}
\end{equation} 
in the Einstein frame, 
then from (\ref{scalefactoreinst}) 
it becomes clear that an order one negative 
value of $q$ is obtained. 

The important point, however, is that 
this is compatible with large enough times $t_E$ (in string units) 
for 
\begin{equation} 
|C_5|e^{-5s_{02}} \gg 1~. 
\label{largetwe}
\end{equation} 
as becomes clear from the definition of $\gamma$ (\ref{defA}). 
This condition can be guaranteed
{\it either} for small radii of the five of the extra dimensions 
{\it or} for a large value of the flux $|C_5|$ of the five-form 
of the type-$0$ string. Notice that the relatively large extra dimension,
in the direction of the flux, 
$s_{01}$, decouples from this condition, thus 
allowing for the possibility of effective five-dimensional models 
with large uncompactified fifth dimension.  

Notice that in the regime (\ref{condition}) of Einstein-frame times 
the Hubble parameter
and the cosmological constant will continue to be compatible
with the current observations, and in fact to depend on $\gamma \sim t_E^{-1}$ 
as in their large $\gamma t_E$ regime given above (\ref{hubble2}),(\ref{cosmoconst}). This was 
expected
from simple dimensional analysis and can be confirmed
by a detailed analysis of the respective formulae 
obtained from (\ref{scalefactoreinst}).

We next look at the equation of state 
of our type-0 string Universe. 
As discussed in \cite{dgmpp}, our situation is a 
quintessence like case, with the dilaton playing the r\^ole of the quintessence
field~\cite{emnsmatrix,dgmpp}. 
Hence the equation of state
for our type-$0$ string Universe reads~\cite{carroll}:
\begin{equation}\label{eqnstate} 
          w_\Phi = \frac{p_\Phi}{\rho_\Phi}=\frac{\frac{1}{2}({\dot \Phi})^2 - V(\Phi)}
{\frac{1}{2}({\dot \Phi})^2 + V(\Phi)}
\end{equation}
where $p_\Phi$ is the pressure and $\rho_\Phi$ is the energy density, 
and $V(\Phi)$ is the effective potential for the dilaton, which in our case
is provided by the central-charge deficit term. 
Here the dot denotes Einstein-frame differentiation. 

In the Einstein frame the 
potential $V(\Phi)$  is given by $\Lambda _E $ in (\ref{cosmoconst}).
In the limit $Q \to q_0$, which has been 
argued to characterize the present era
to a good approximation, the present era 
$V(\Phi)$ is then of order                             
$(q_0^2/2F_1^2)t_E^{-2}$, where we recall 
that $q_0/F_1$ is
given by (\ref{ci}).

In the Einstein frame
the exact normalization of the dilaton field is 
$\Phi _E = {\rm const} -{\rm ln}t_E $.
Combining this result with (\ref{ci}), we then 
obtain for the present era (\ref{condition}): 
\begin{equation}\label{dilpotkin}  
\frac{1}{2}{\dot \Phi}^2 \sim \frac{1}{2t_E^2}, \qquad V(\Phi) 
\sim \frac{6.56}{2}\frac{1}{t_E^2}
\end{equation} 
This 
implies an equation of state (\ref{eqnstate}): 
\begin{equation} 
w_\Phi (t_E \gg 1) \simeq -0.74 
\label{eqnstatedil}
\end{equation}
for (large) times $t_E$ in string units corresponding to the present era
(\ref{condition}).

We should compare this result 
with the case of an ``honest'' cosmological {\it constant} 
situation, as the one fitting the data in \cite{evidomegal},
which yields $w=-1$, or 
with the standard
scenario of 
a {\it slow-varying quintessence
field} $\varphi$, 
${\dot \varphi}^2 \ll V(\varphi)$, which again yields 
$w_\Phi \simeq -1$ by means of (\ref{eqnstate}). 
The reader should recall that, in our case, 
the result (\ref{eqnstatedil}) occurs because 
the relative magnitude (\ref{dilpotkin}) of the 
dilaton (quintessence) potential versus
that of its kinetic 
energy is fixed by conformal invariance (\ref{ci}). 
It would be interesting to 
see whether the inclusion of 
ordinary matter (attached on the 3-branes of the type-0 string) 
changes these results significantly.
This is left for future work.

If the present era of the Universe is therefore 
defined by (\ref{condition}),
then one obtains {\it compatibility} of the above results 
with the {\it current astrophysical 
observations}~\cite{evidomegal,evidenceflat}:

\begin{equation} \label{phenomen}
H(t_E) \sim \frac{1}{t_E}~, \qquad \Lambda (t_E) 
\sim \frac{{\cal O}(1-10)}{t_E^2} > 0, \qquad q(t_E) \sim -|{\cal O}(1)| < 0 
\end{equation}

It is amusing to see here that in order to get an order-one 
deceleration parameter (\ref{phenomen}), as the present phenomenology
suggests~\cite{evidomegal}, 
it appears necessary to have 
the ratio of the flux $C_5$ of the five-form of the type 0 string
over the volume of the (five) transverse dimensions $e^{5\sigma_{02}}$ 
very large. If one insists on keeping 
the flux $C_5$ of order one, something, however, which is {\it
not necessary}, 
then 
this result implies 
{\it transplanckian} 
(smaller than Planck length) sizes of at least five of 
the extra 
dimensions, so that the volume $V_6 \ll 1$ in string units.

Such small volumes in turn require 
bulk string mass scales much larger than the 
Planck mass scale on the four-dimensional brane world. 
To see this recall that the gravitational part of 
the bulk ten-dimensional target-space effective 
action (\ref{actiontype0}) reads
after compactification in the Einstein frame:
\begin{equation}
S \sim \frac{1}{g_{10}^2}M_s^{8} V_6 \int d^4x \sqrt{g^{(4)}} R_E ^{(4)} 
+ \dots 
\label{fourgrav}
\end{equation} 
where $g_{10} \sim e^{\langle \Phi \rangle}$ is the ten-dimensional
string coupling, which for our purposes here is assumed weak $g_{10}  < 1$. 
The overall coefficient in front of the four-dimensional Einstein term
defines the square of the four dimensional Planck scale $M_P$:
\begin{equation}
M_P^2 = \frac{1}{g_{10}^2}M_s^8 V_6
\end{equation}
Therefore, for small $V_6 \ll 1$ in string units, i.e. 
$M_s^6 V_6 \ll 1$, one obtains that the string mass scale
should be much larger  than the four-dimensional Planck mass scale:  
\begin{equation}
\label{twoscales}
M_P \ll M_s~. 
\end{equation} 
Notice that, in the modern viewpoint that the string bulk scale is a 
free parameter in string/M-theory, such relations are allowed. 
This is a curious  feature of our construction, and certainly
requires further study; in particular, one should investigate
possible connections between 
our approach and that of \cite{branden}, where transplanckian
modes have been argued to play an important r\^ole for inflation.

Notice that the above results have been obtained without the inclusion 
of ordinary matter. 
In the type-$0$ string/brane scenario, ordinary matter may be assumed
attached to the brane, and hence purely four dimensional. 
The latter is going to resist the deceleration 
of the universe, according to standard arguments, but it is not expected
to change the order of magnitude of the above quantities.
In this sense one may obtain the observed `coincidence situation' of the 
present era, where the matter and `dark energy' contributions 
are roughly of the same order of magnitude~\cite{carroll}. 
In our scenario it is the time dependence of both `dark energy' and
`matter contribution', in conjunction 
with the value the time $t_E$ has at present, 
roughly $t_E \sim 10^{60}M_P^{-1}$, 
that is held responsible for this coincidence
situation. 

%%%%%%%%%%%%%%%%%%%%%%%%%%%%%%%%%%%%%%%%%%%%%%
\begin{figure}[h]
\centering
\includegraphics[scale=0.5]{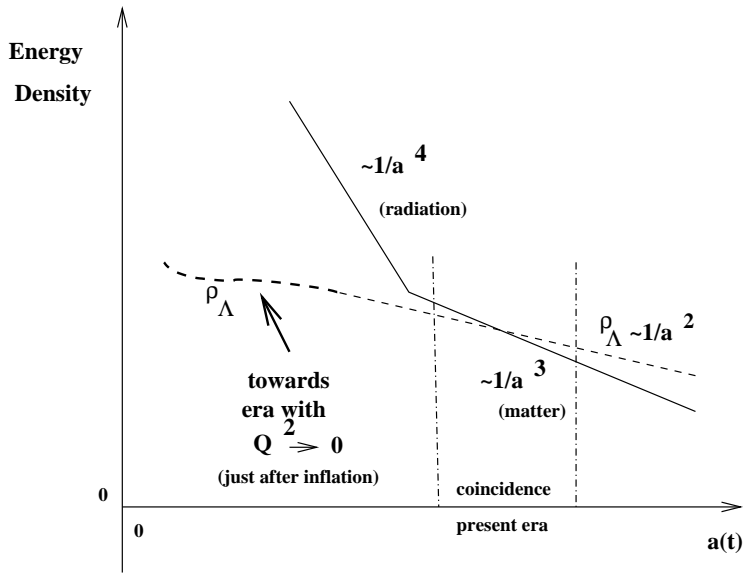}
%%\centerline{\hbox{\psfig{figure1b=figure1b.eps,height=10cm}}}
\caption {The evolution of the energy densities of matter, radiation
and of the quintessence field (dilaton) vs. the scale factor 
of the Universe  
in the Einstein frame. At early stages the energy density of the 
quintessence field decreases significantly, as compared with the 
rest, and the coincidence situation is lost. This is due to the 
behaviour of the central charge deficit of the model, shown 
in figure \ref{centralcharge:fig},
which dives in to zero for a short period 
immediately after inflation.} 
\label{energydensity:fig}
\end{figure}
%%%%%%%%%%%%%%%%%%%%%%%%%%%%%%%%%%%%%%%%%%%%%%

Indeed, from fig. \ref{centralcharge:fig} 
we observe that at relatively early times (after inflation) 
there is a period where the central charge deficit $Q^2$, and hence the 
potential energy $V(\Phi)$ of the quintessence `tracking' dilaton field $\Phi$,
vanishes. The dilaton equation, which guarantees conformal invariance
of the Liouville dressed theory~\cite{dgmpp}, implies near this point that
$(d\Phi/dt_\sigma )^2 \sim d^2 \Phi/dt_\sigma ^2$, where 
$t_\sigma =-{\rm ln}t_E$ 
is the $\sigma$-model time. This means that the dilaton field 
kinetic energy density, and hence its total energy density $\rho_\Lambda$,   
scales logarithmically with the Einstein time in this (short) time region
$\rho_\Lambda \propto 1/({\rm ln}t_E +{\rm const})^2 $.  
Shortly after this point, as the time elapses
the central charge increases significantly in such a way that 
$d^2\Phi/dt^2 < 0$, and eventually the dilaton reaches its linear 
equilibrium configuration in string frame. This behaviour is dictated
by conformal invariance of the underlying $\sigma$-model. 
Taking into account (\ref{dilaton}),(\ref{sigmatime}), 
this implies that, as one goes backwards in time, 
starting from  the present era, the energy 
density of $\Phi$, $\rho_\Phi =\frac{1}{2}({\dot \Phi})^2 +V(\Phi)$, 
becomes significantly smaller than the energy density of matter (or radiation),
which increase as the time goes backward,
scaling with the scale factor like $a_E^{-3}$ (or $a_E^{-4}$).
Thus in our model 
the tracking (coincidence) of the matter energy density by that of the 
quintessence dilaton field 
is a feature only of the present era, which is a welcome feature 
phenomenologically.  The situation is summarized in figure
\ref{energydensity:fig}, where we plot (in a qualitative manner) 
the energy densities of radiation, 
matter and of the quintessence field (dilaton),  
$\rho_\Lambda$, vs. the scale factor $a_E(t_E)$ 
of the Robertson-Walker non-critical string Universe
(in the Einstein frame). The plot, which is not to scale,  
is based on the (qualitative) behaviour
of the $Q^2$, shown in  
figure \ref{centralcharge:fig}, and the above discussion.

As the time $t_E$ elapses, the matter contribution will become subdominant,
as scaling like $a_E^{-3}$. For very large
times $t_E$ in the far future, as we have seen above, 
the dominant contributions will be the ones 
due to the 
non-constant in time 
`dark energy component' 
$\Lambda (t_E) \sim a_E^{-2}$, which asymptotes to zero,
as the system reaches its equilibrium value.  
This makes a quantitative difference 
in scaling as compared with the standard Robertson-Walker scenario
with a constant vacuum energy (c.f. (\ref{conventional})).

\section{Conclusions}

In this note we have discussed the cosmological evolution of the present
era deceleration parameter of a non-critical type-$0$ string Universe,
and we have argued that one can get compatibility with current
astrophysical observations. We have been able to fit the data with a 
quintessence-like non-critical 
string Universe, which has a present-era negative 
deceleration 
parameter $q$ of order one, in agreement with the supernova Ia observations.
The dark energy of our string Universe, however, 
is not constant, but relaxes to zero 
asymptotically, 
in a way compatible with the current value of the dark energy,
explaining 
the observed `coincidence' in the order of magnitudes between matter
and dark energy components as a matter of `chance' (this is like
an anthropic principle situation: we have been `lucky' to witness
this event, being in the right `place' at the `right time').

We do not claim here that our crude string model is a physical
model for the Universe, but we find it interesting that at
least to a first approximation the phenomenology of the model
seems to match the data. 
An important r\^ole for obtaining  an order-one 
value for the deceleration parameter 
is played by the relative value of the flux of the five-form field of the
type-$0$ string to the volume of the small extra dimensions. 
Interestingly enough the value of the deceleration parameter
is independent of the large bulk dimension, along the direction
of which lies the flux of the five form of the type-$0$ string. 
This leaves room for accommodating effective 
five-dimensional scenaria, with large uncompactified fifth dimension.

We therefore 
consider these considerations very interesting, and certainly worthy of 
further investigations. It is our belief that these characteristics 
extend beyond the specific model of type-$0$ string theory
studied here.
In fact we think that 
such a behaviour  may characterize 
a large class of non-critical (non-equilibrium)
string Universes, 
relaxing to their critical situation asymptotically in 
time. It will certainly be interesting to study properly the phenomenology
of such models, by taking into account 
the detailed form of string matter,
attached to the brane worlds, including fermionic excitations, 
and see whether interesting predictions
can be made, in the cosmological sense, that could differentiate among the various models. We hope to come back to such a study in the near future.

\section*{Acknowledgements}

G.A.D and B.C.G. would like to acknowledge partial
financial support from the Athens University special account for
research. N.E.M. wishes to thank Subir Sarkar for discussions. 
E.P. wishes to thank the Physics Department
of King's College London for the warm hospitality 
during the last stages of this work.

\end{document}